\documentclass[aps,pra,reprint,showpacs,preprintnumbers,amsmath,amssymb,superscriptaddress]{revtex4-2}

\usepackage{graphicx}
\usepackage{dcolumn}
\usepackage{bm}
\usepackage{color}
\usepackage{bbm}
\usepackage{units}
\usepackage{array}
\usepackage{tabularx}
\usepackage{multirow}
\usepackage{amsmath,amssymb,amsfonts}
\usepackage{pifont}
\usepackage[colorlinks=true,citecolor=blue,linkcolor=blue,urlcolor=blue]{hyperref}

\begin{document}

\title{Efficiently Generation of Cluster States via Time-Delayed Feedback \\ in Matrix Representation}

\author{Jia-Jin Zou}
\affiliation{School of Physics, Sun Yat-sen University, Guangzhou 510275, China}
\author{Jian-Wei Qin}
\affiliation{School of Physics and Astronomy, Shanghai Jiao Tong University, Shanghai 200240, China}
\author{Franco Nori}
\affiliation{Center for Quantum Computing, RIKEN, Wako-shi, Saitama 351-0198, Japan}
\affiliation{Department of Physics, University of Michigan, Ann Arbor, Michigan 48109-1040, USA}
\author{Ze-Liang Xiang}
\email{xiangzliang@mail.sysu.edu.cn}
\affiliation{School of Physics, Sun Yat-sen University, Guangzhou 510275, China}
\affiliation{State Key Laboratory of Optoelectronic Materials and Technologies, Sun Yat-sen University, Guangzhou 510275, China}
\date{\today}

\begin{abstract}
Cluster states, as highly entangled multi-qubit states, are widely used as essential resources for quantum communication and quantum computing. However, due to the diverse requirements of applications for cluster states with specific entanglement structures, a universal generation protocol is still lacking. Here we develop a matrix representation according to the characteristics of time-delayed feedback (TDF) and propose a protocol for generating arbitrary cluster states with multiple TDFs. The matrix representation also allows us to optimize the generation process to reduce TDF usage, thus improving efficiency. In particular, we demonstrate a tree-cluster-state generation process that requires only one TDF. Moreover, accounting for the critical loss mechanisms and imperfections in our protocol, we discuss the additional losses caused by multiple TDFs and evaluate the fidelity of the resulting cluster states.
\end{abstract}
\maketitle

\section{Introduction}

Cluster states were first proposed as resources for measurement-based quantum computing (MBQC)~\cite{PRL_raussendorf_2001, PRA_raussendorf_2003, NatPhys_briegel_2009, PRA_gross_2007}. Unlike quantum circuit architectures, the processes of information transfer and manipulation are realized by quantum teleportation between pairs of entangled qubits in the cluster state~\cite{NewJ.Phys_van_2007, J.PhysB_barz_2015, PRL_nielsen_2005}. Hence, the circuit architecture's coherent unitary evolution (quantum gates) can be achieved by performing sequences of adaptive measurements~\cite{J.PhysB_barz_2015,Nat.Phys_larsen_2021,PRA_Yang_2022}. Concretely, the potential applications of cluster states depend on their entanglement structures. MBQC typically requires cluster states with lattice structures~\cite{PRL_van_2006,NewJ.Phys_raussendorf_2007},  while complete-like and tree structures are commonly employed in quantum communication or quantum error correction~\cite{PRA_pant_2017,PRX_buterakos_2017,PRA_wallnofer_2017, NatCom_azuma_2015_repeater,NPJ-QI_niu_2023,PRA_zwerger_2012,ApplPhysB_zwerger_2016,Nat.Com._Bonilla_2021},

To date, intense efforts have been devoted to generating cluster states, with one-dimensional (1D) chain-like cluster states being well studied and realized~\cite{PRL_Nori_2006,PRL_lindner_2009,Science_schwartz_2016,NatCom_istrati_2020,PRA_tiurev_2022}. In order to obtain more functional cluster states, many methods have been tried to generate more complex entanglement structures. On the one hand, higher-dimensional lattice cluster states used for MBQC have been generated in continuous-variable systems~\cite{Science_asavanant_2019,Science_larsen_2019,Nature_thomas_2022,Nat.Pho._roh_2025_3DLCS}, and other methods of encoding at different degrees of freedom have also been studied~\cite{NPJ_wang_2018,Nat.Pho._cogan_2023,AAPPS_Huang_2023,Nature_cao_2023,Nat.Pho._lib_2024}. On the other hand, the exploration of the generation of tree cluster states (TCSs) and complete-like cluster states (CCSs) has so far been confined to theoretical protocols~\cite{PRX_buterakos_2017,PRX_borregaard_2020}. 
Recently, the concept of quantum feedback~\cite{Phys.Rep._Nori_2017} was introduced to create additional entanglement in 1D cluster states~\cite{PRL_Arne_2015,arxiv_cheng_2024}. One of the typical protocols is to generate 2D lattice cluster states via a time-delayed feedback (TDF) in a waveguide~\cite{PNAS_pichler_2017}, which has been demonstrated in experiments~\cite{Nat.Phys._ferreira_2024}. Further, some progress has been made in reusing TDF or introducing multiple TDFs to produce more complex cluster states~\cite{PRL_zhan_2020,PRA_shi_2021}.

In contrast to the site-to-site interaction in previous protocols, the TDF-induced interaction usually works for a series of photons uniformly arranged in the time domain. Inspired by the reusability of each TDF for generating specific entanglement structures, one can explore a more efficient way of utilizing this resource to produce the required cluster states. However, existing cluster state representations fail to capture the regularity of qubit arrangements, thus necessitating a new representation to reveal more features of the entangled structures of cluster states.

In this work, we develop a matrix representation of cluster states that suits the characteristics of TDFs, which can classify the entanglement in a cluster state according to certain indices, such as the time interval between two entangled qubits. Based on this matrix representation, we propose a universal protocol that can generate arbitrary cluster states via multiple TDFs. By introducing virtual nodes to manipulate the excited time of each qubit, the regularity of entanglement structures can be extracted. This allows us to optimize the generation process by minimizing TDF usage. Particularly, for binary tree-like cluster states (${\rm{TCS}}_{2,d}$), the usage of TDFs can be reduced from exponential $\mathcal{O}(2^d)$ to linear $\mathcal{O}(d)$ in the tree depth $d$. For $d\leq5$, ${\rm{TCS}}_{2,d}$ can even be generated using a single-TDF system. Moreover, we consider amplitude damping as an additional loss in multiple-TDF systems and calculate the influence of amplitude damping and the imperfection of quantum gates on the fidelity of the resulting cluster states, confirming its resource efficiency and resilience to losses.

To facilitate reading, we list the several abbreviations used in this article in Table~\ref{Abb}. 

\begin{table}[t]
    \setlength{\tabcolsep}{0.2cm}
    \renewcommand{\arraystretch}{1.3}
    \centering
    \begin{tabular}{>{\centering\arraybackslash}m{2.2cm}|>{\centering\arraybackslash}m{5.4cm}}
    \hline
        \textbf{Abbreviation} & \textbf{Definition} \\ \hline
        MBQC & Measure-based quantum computing \\ 
        TCS & Tree cluster state \\ 
        CCS & Complete-like cluster state \\ 
        TDF & Time-delayed feedback \\ 
        CZ  & Controlled-Z \\ \hline
    \end{tabular}
    \caption{List of Abbreviations}
    \label{Abb}
\end{table}


\section{Matrix representation of cluster states}

A quantum cluster state can be uniquely described by a graph $G=(V,E)$ consisting of \textit{nodes} $V$ and \textit{edges} $E$, where nodes and edges correspond to qubits and entanglement between qubits, respectively. This is called the graph representation and can accurately describe the entanglement structure of the cluster state. Alternatively, a cluster state $|\psi_C\rangle$ can also be represented as a series of gate operations acting on a given initial state~\cite{QuanSci_morley_2017} 
\begin{equation}
    |\psi_C\rangle\equiv\left[\prod_{(i,j)\in E} \hat{C}^{Z}_{i,j}\right]\otimes_{i'\in V}|+\rangle,
    \label{eq1}
\end{equation}
where $\hat{C}^{Z}_{i,j}$ denote the controlled-Z (CZ) gates between the $i$-th and $j$-th qubits, and $|+\rangle=(|0\rangle+|1\rangle)/\sqrt{2}$. The information about the entanglement structure is not intuitive within the gate representation, but the generation process of cluster states is clearly illustrated. Accordingly, the generation of the cluster state in Eq.~(\ref{eq1}) can be divided into two steps: 
\begin{enumerate}
    \item[(1)] preparing all qubits into the initial state $|+\rangle$; 
    \item[(2)] applying CZ gates between qubits as needed. 
\end{enumerate}

\begin{figure}[t]
    \centering
    \includegraphics[width=1\linewidth]{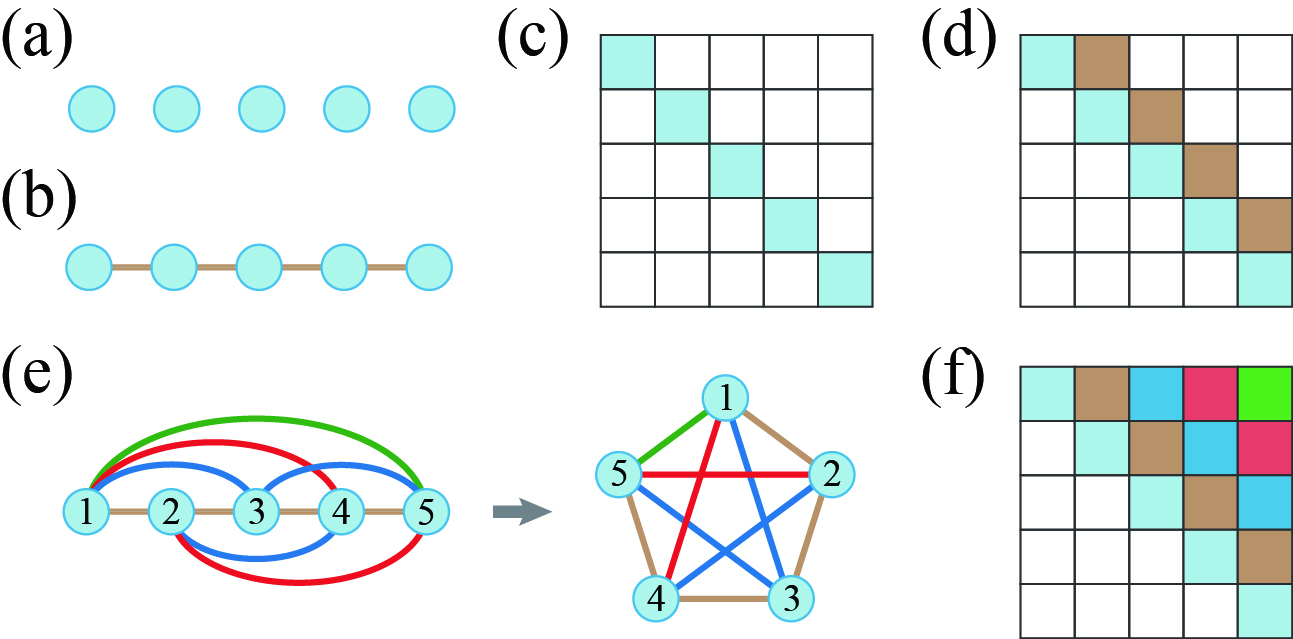}
    \caption{Diagrams for different states and visualization of their matrix representations. (a) Five individual qubits and (b) a 5-qubit 1D cluster state, whose matrix representations are drawn in (c) and (d), respectively. (e) The transformation from a 1D cluster state to a 5-qubit complete-like cluster state $({\rm{CCS}}_{5})$, where different edge colors indicate different kinds of entanglement. (f) The matrix of CCS$_{5}$. The blue diagonal units mean excitations of qubits, and the upper triangular units in different super diagonals are denoted with different colors.}
    \label{FIG1}
\end{figure}

In order to extract the commonality of quantum gates used for generating cluster states, we employ a block matrix ${\bf M}_{2N\times N}$, with each element being a vector, to describe the cluster state. Here, $N$ is the number of qubits available to form the cluster state. Such a matrix can be regarded as a square matrix, and the elements (vectors) located in the diagonal, upper, and lower triangle parts are used to denote different components of the cluster state. Without loss of generality, the ${\bf M}_{2N\times N}$ matrix of the cluster state can be written in the form
\begin{equation}
    {\bf M}_{2N\times N}=
    \begin{pmatrix}
    M_{1} & M_{1,2} & \cdots & M_{1,n} \\
    0 & M_{2} & \cdots & M_{2,n} \\
    \vdots & \vdots & \ddots & \vdots \\
    0 & 0 & \cdots & M_{n}
    \end{pmatrix}.
\end{equation}
The nonzero elements are located at the diagonal part $M_{i}$ and the upper triangle part $M_{i,j}$ ($i<j$), each of them being in the state $|0\rangle=(0,1)^T$ or $|1\rangle=(1,0)^T$.

The state of $M_i$ (corresponding to the excitation of the $i$-th qubit, i.e., $|0\rangle$ and $|1\rangle$) describes unexcited and excited states, respectively. Only two-qubit entanglement is considered in the cluster state, which completely determines the upper triangular elements $M_{i,j},(j>i)$. If there is entanglement between the $i$-th and the $j$-th qubits, the corresponding element should be in the state $M_{i,j}=|1\rangle$. Conversely, if $M_{i,j}=|0\rangle$, there is no correlation between the two qubits. Moreover, the lower triangular elements of $\bf M$ are all zero vectors, denoted by $0$.

In order to describe the generation process of a cluster state, we decompose the matrix representation of the target state into the form of the initial matrix ${\bf M}_0$ and the sequential action of the operators. Initially, all nonzero matrix elements in ${\bf M}_0$ are set to $|0\rangle$, which means no qubits have been excited. Then we introduce the excitation operator $X_{i}$ and entanglement operation $E_{i,j}$, which can turn the state of the diagonal elements $M_i$ and upper triangular elements $M_{i,j}$ of the matrix from $|0\rangle$ to $|1\rangle$, respectively.

\subsection{One-dimensional Cluster States}

Before discussing the physical meaning of the subscript of matrix elements, we first show the evolution process of the matrix of an $N$-qubit 1D cluster state from ${\bf M}_0$. In a 1D cluster state, every qubit will be excited and entangled to its two adjacent qubits, and this process can be written in the form of operators' sequential action on the initial matrix ${\bf M}_0$:
\begin{equation}
    {\bf M}_{1D,N}=\prod_{i=2}^N E_{i-1,i} \prod_{i=1}^{N} X_{i} {\bf M}_0.
    \label{eq3}
\end{equation}

Obviously, the entanglement operators $\left\{ E_{i-1,i} \right\}$ used in this process have the same form. If we focus on a photonic cluster state, the subscript can be considered as the time order in which the photon was emitted under a fixed emission interval $\tau_{0}$. Then a TDF with a specific delay factor $\alpha$ will induce a series of entanglement operators $\left\{ E_{i-\alpha,i} \right\}$; which sometimes can be simply represented as $E_{i,\alpha}$, where $i$ indicates the target qubit of the CZ gate and $\alpha$ indicates the interval between the target qubit and the control qubit.

Further, when $N=5$, the resulting matrix is 
\begin{equation}
    {\bf M}_{1D,5}=
    \begin{pmatrix}
    |1\rangle & |1\rangle & |0\rangle & |0\rangle & |0\rangle \\
    0 & |1\rangle & |1\rangle & |0\rangle & |0\rangle \\
    0 & 0 & |1\rangle & |1\rangle & |0\rangle\\
    0 & 0 & 0 & |1\rangle & |1\rangle \\
    0 & 0 & 0 & 0 & |1\rangle
    \end{pmatrix}.
\end{equation}
The diagrams for the matrix representations of a series of five individual qubits and a 5-qubit 1D cluster state are shown in Figs.~\ref{FIG1}(a)-(d), and the corresponding visualization graphs can help us to extract the essential features of the matrix representation.

In ${\bf M}_{1D,5}$, we find that only the diagonal elements, and elements in the first diagonal of the upper triangle matrix (denoted as the first super diagonal) are in the state $|1\rangle$, while other upper triangular elements are all in the state $|0\rangle$. According to the number of super diagonals including $|1\rangle$-state elements in the matrix representation, we are able to label how many kinds of entanglement the cluster state contains. Thus, we can determine how many TDFs are required to generate this cluster state. It is clear that a 1D cluster state contains only one kind of entanglement, and its generation has no need of TDF.

\subsection{Complete-like Cluster States}

Another widely used cluster state has a complete-graph structure, called a complete-like cluster state (CCS). The matrix used to represent CCS can be generated via the process:
\begin{equation}
   {\bf M}_{{\rm CCS}_N}=\prod_{\alpha=1}^{N-1} \left(\prod_{i=1}^{N} E_{i,\alpha}\right) \prod_{i=1}^{N} X_{i}{\bf M}_0.
\end{equation}
For instance, the matrix of a 5-qubit CCS is
\begin{equation}
    {\bf M}_{\rm CCS_5}=
    \begin{pmatrix}
    |1\rangle & |1\rangle & |1\rangle & |1\rangle & |1\rangle \\
    0 & |1\rangle & |1\rangle & |1\rangle & |1\rangle \\
    0 & 0 & |1\rangle & |1\rangle & |1\rangle\\
    0 & 0 & 0 & |1\rangle & |1\rangle \\
    0 & 0 & 0 & 0 & |1\rangle
    \end{pmatrix}.
\end{equation}
All elements in ${\bf M}_{{\rm CCS}_N}$ are in the state $|1\rangle$, except for the lower triangle part, which implies that ${\rm CCS}_{N}$ is the maximally entangled state including $(N-1)$ kinds of entanglement. 

Using column vectors as matrix elements is a necessary condition to establish the representation matrix $\bf{M}$ for a cluster state. But in the subsequent content, the $\bf{M}$ can be simplified to a distribution matrix $\bf{D}$ that describes the distribution of the $|1\rangle$-state elements, which is enough for the subsequent analysis of cluster states.

Mathematically, the distribution matrix $\bf{D}$ can be obtained by projecting $\bf{M}$ onto the $(1,0)$ basis. The projection matrix ${\bf P}_N= {\bf{I}}_N \otimes (1,0)$ and ${\bf{D}}={\bf{P}}_N \bf{M}$, where $N$ is the dimension of the state. For example, the distribution matrix for ${\bf M}_{1D,5}$ is
\begin{equation}
    {\bf D}_{1D,5}={\bf P}_{5}{\bf M}_{1D,5}=
        \begin{pmatrix}
    1 & 1 & 0 & 0 & 0\\
    0 & 1 & 1 & 0 & 0 \\
    0 & 0 & 1 & 1 & 0\\
    0 & 0 & 0 & 1 & 1 \\
    0 & 0 & 0 & 0 & 1
        \end{pmatrix}.
\end{equation}

\section{Protocol}
\label{protocol}

Based on the matrix representation, we can determine how many TDFs are required to generate a cluster state. Now, we should provide a protocol that allows us to use multiple TDFs to generate cluster states efficiently.

\begin{figure}
    \centering
    \includegraphics[width=1\linewidth]{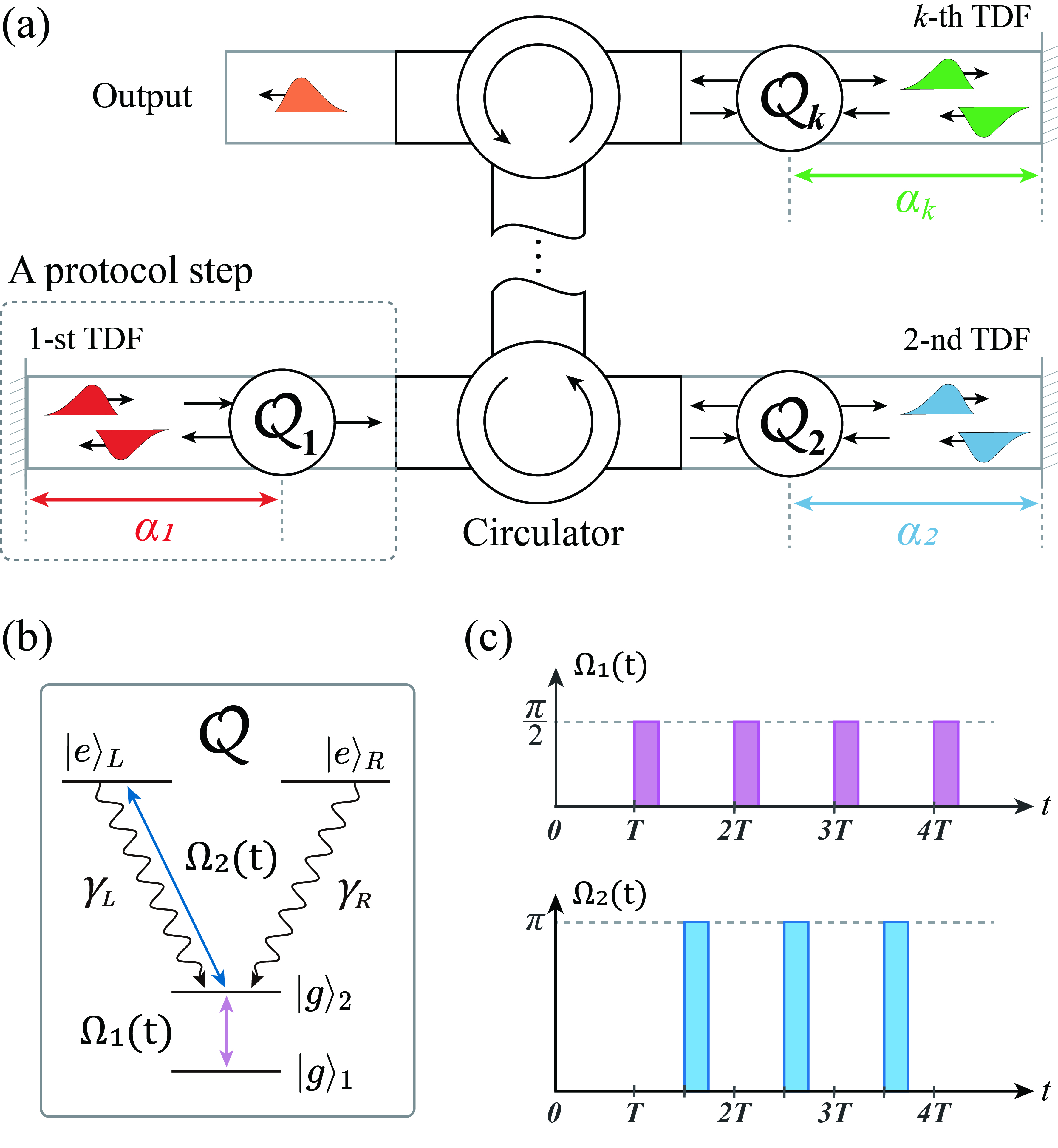}
    \caption{Schematic diagram of the universal protocol that generates cluster states with multiple TDFs. (a) Each block consists of an emitter coupled to the time-delayed feedback, which establishes a TDF. The connection between different blocks is achieved by the circulator. (b) The energy level of each emitter in our protocol, the classical field $\Omega_{1}(t)$ can manipulates the transition between $|g\rangle_{1}$ and $|g\rangle_{2}$, the pulse $\Omega_{2}(t)$ can excite the emitter from $|g\rangle_{2}$ to $|e\rangle_{L}$. For generating a 1D cluster state, the pulse sequence as depicted in (c) must be applied on the emitter.}
    \label{FIG2}
\end{figure}

The schematic diagram of our protocol that consists of multiple TDFs is shown in Fig.~\ref{FIG2}(a). The dashed box of Fig.~\ref{FIG2}(a) emphasizes a fundamental protocol step, which consists of a quantum emitter ($\mathcal{Q}$) coupled to a 1D waveguide and a mirror, the distance between $\mathcal{Q}$ and the mirror will determine the delay factor $\alpha$ of the TDF. 

The energy level of $\mathcal{Q}$ is depicted in Fig.~\ref{FIG2}(b), showing a $V$-type energy level (two excited states $|e\rangle_{R}$, $|e\rangle_{L}$ and a ground state $|g\rangle_{1}$) and a auxiliary level. The two metastable states $|g\rangle_{1} \equiv |0\rangle_{\mathcal{Q}}$ and $|g\rangle_{2} \equiv |1\rangle_{\mathcal{Q}}$ of $\mathcal{Q}$ can be coherently manipulated by a classic field $\Omega_{1}(t)$. At the same time, the emitter can be excited from $|1\rangle_{\mathcal{Q}}$ to $|e\rangle_{R}$ using a laser with Rabi frequency $\Omega_{2}(t)$. Following each excitation, the emitter will decay to $|g\rangle_{2}$ and emit a photon into the waveguide.

In our protocol, each photon is encoded via the absence $|0\rangle_{i}$ or the presence $|1\rangle_{i}$ in the $i$-th pulse, which can be considered as a photon number basis. We stress that under the modulation of a circular polarization electric field, the transition from $|g\rangle_{2}$ to $|e\rangle_{R}$ ($|e\rangle_{L}$) will couple with the right-propagating (left-propagating) photon, respectively~\cite{Nature_lodahl_2017_chiral,Nat.Pho._Nori_2015,PRL_Nori_2008}.

The generation of a photonic cluster state begins with generating a 1D cluster state on the first block. The initial state of $\mathcal{Q}_{1}$ is set to $|g\rangle_{1}$. Then we apply a rapid $\pi/2$-pulse to excite $\mathcal{Q}_{1}$ into the superposition state $|+\rangle$, followed by a $\pi$-pulse on the $|g\rangle_{2}\rightarrow|e\rangle_{L}$ transition. The subsequent spontaneous radiation from $|e\rangle_{L}$ will emit a left-propagating photon, at that time the state of the whole system is $|0\rangle_{\mathcal{Q}}|0\rangle_{1}+|1\rangle_{\mathcal{Q}}|1\rangle_{1}$. Repeating this pulse sequence for $N$ times in a period $T=\tau_{0}$ (as depicted in Fig.~\ref{FIG2}(c)), an $N$-qubit 1D cluster state can be generated.

After the $i$-th photon leaves the emitter $\mathcal{Q}_{1}$, it will be reflected by the mirror and return to $\mathcal{Q}_{1}$ after a time $T'=\alpha_{1}\times\tau_{0}$. Under the periodic drive of the pulse sequence, the $i$-th photon will reach the emitter just before the emission of the $(i+\alpha_{1})$-th photon. At that time, if $\mathcal{Q}_{1}$ is in $|1\rangle_{\mathcal{Q}}$, it will absorb the returning left-propagating photon and transit to $|e\rangle_{R}$. Then the following decay, from $|e\rangle_{R}$ to $|1\rangle_{\mathcal{Q}}$, will emit a photon with a $\pi$-phase shift, which is equal to a $\pi$-phase scattering without reflection of photons~\cite{Nature_lodahl_2017_chiral,Nature_tiecke_2014}. In contrast, if the emitter is in $|0\rangle_{\mathcal{Q}}$ at that moment, there is no interaction between the returning photon and the emitter. Combining with the subsequent emission process of the $(i+\alpha_{1})$-th photon, we can rewrite this interaction in the form of a CZ gate between the $i$-th and $(i+\alpha_{1})$-th photons:
\begin{equation}
\hat{C}^{Z}_{i+\alpha_{1},i}=|0\rangle_{i+\alpha_{1}}\langle0|\otimes I_{i} + |1\rangle_{i+\alpha_{1}}\langle1|\otimes \sigma_{i}^{z}.
\end{equation}
In short, the TDF has a function that entangles photons with a specific time interval, just as a series of entanglement operators $E_{i,\alpha_{1}}$.

Then, photons are guided into the next block with the help of a photonic circulator. In the second block, which is used to introduce a new TDF, the process will be a little different. First, we should change the polarization direction of the applied electric field, and thus, the $\pi$-phase scattering occurs on the left-propagating photon. At the beginning of each cycle, a $\pi$-pulse is applied on $\mathcal{Q}_{2}$ to prepare it to the state $|1\rangle_{\mathcal{Q}}$, but the absorption and emission have no effect on the state of the right-propagating photon.

The $i$-th photon will be reflected and reach $\mathcal{Q}_{2}$ again after a time delay $\tau_{2}=\alpha_{2}\times\tau_{0}$, which is independent of the first TDF. In our protocol, the right-propagating $(i+\alpha_{2})$-th photon will arrive at the position of $\mathcal{Q}_{2}$ just before the reflected left-propagating $i$-th photon. If the $(i+\alpha_{2})$-th photon is in $|1\rangle_{i+\alpha_{2}}$, the emitter $\mathcal{Q}_{2}$ will be excited to $|e\rangle_{R}$ by absorbing the photon, then the reflected $i$-th photon will not interact with $\mathcal{Q}_{2}$. On the contrary, if the $(i+\alpha_{2})$-th photon is in $|0\rangle_{i+\alpha_{2}}$, $\mathcal{Q}_{2}$ will remain in the state $|1\rangle_{\mathcal{Q}}$. Due to the chiral coupling, the returning $i$-th photon will pick up a $\pi$-phase shift without reflection. This process is also equivalent to a CZ gate between the $i$-th and $(i+\alpha_{2})$-th photons,
\begin{equation}
\hat{C}^{Z}_{i+\alpha_{2},i}=|0\rangle_{i+\alpha_{2}}\langle0|\otimes I_{i} + |1\rangle_{i+\alpha_{2}} \langle1|\otimes \sigma_{i}^{z},
\end{equation}
which can also be represented as $E_{i,\alpha_{2}}$, and the two delay factors are independent of each other.

The same process is repeated in other blocks, resulting in any number of TDFs with different delay factors. Integrating with the matrix representation, our protocol provides a feasible way to generate arbitrary cluster states. Note that there is still a lack of efficient circulator schemes for the single-photon situation, and we present a possible design as a reference in Appendix~\ref{imp}.


\section{Optimization based on matrix representation}
\label{TCS}

\begin{figure*}[t]
\centering
\includegraphics[width=0.95\linewidth]{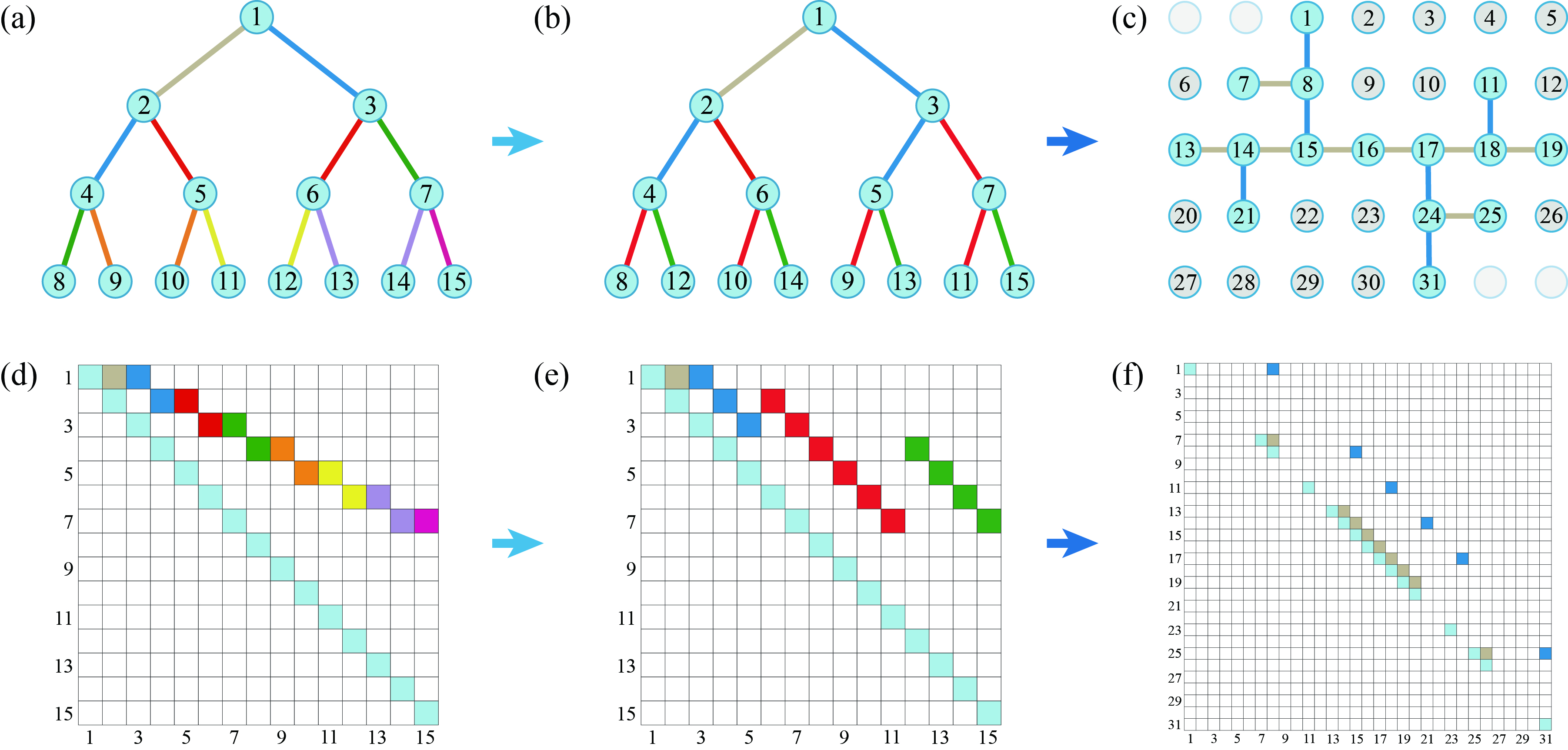}
\caption{Schematic diagram of the optimizing process for a binary tree cluster state (TCS). Each node represents an available photon excitation. Blue nodes mean that photons have been excited, and colored edges represent different kinds of entanglement. (a) is the diagram of the normal way to generate TCS$_{(2,4)}$ from sequentially numbered photons. (d) is the corresponding visualization of its matrix representation, which shows that eight kinds of TDFs are needed for this generation process. (b) and (e) demonstrate the optimization resulting from the adjustment of photon ordering, where only three kinds of TDFs are required. (c) and (f) take virtual nodes into account, TCS$_{(2,4)}$ can be decomposed into two orthogonal directions on a lattice space. Thus, it can be generated more concisely and just needs one additional TDF.}
\label{FIG3}
\end{figure*}

With the universal protocol for generating cluster states via multiple TDFs, reducing TDF usage becomes a priority. The optimization rationale lies in the fact that cluster state properties depend on entanglement topology, not on the number of qubits. When we represent a cluster state with a matrix, different numbering schemes of qubits will give different matrices, which are equivalent in the entanglement structure and behaviors. These ``equivalent'' matrices always lead to different requirements on TDF, which means that one can optimize the generation process by adjusting the representing matrix of the cluster states. 

\subsection{General Optimization of Tree Cluster States}

Let us now consider the tree cluster state (TCS) as an example to illustrate the optimization. In a tree graph, each node connects with $a$ nodes in the next layer, and the number of layers is $d$. Figure~\ref{FIG3}(a) gives a diagram for a simple case (i.e., $a=2, d=4$), where nodes in this tree have been numbered orderly from top to bottom. Then, edges are marked with different colors according to the numbering intervals between connected nodes. The matrix representation of ${\rm{TCS}}_{2,d}$ is shown in Fig.~\ref{FIG3}(d). According to the number of the super-diagonals occupied by colored units, the generation of a ${\rm{TCS}}_{2,d}$ needs seven independent TDFs. Following this numbering method, each TDF is used at most twice, and the number of TDFs needed for generating ${\rm{TCS}}_{2,d}$ can be calculated as $2^{(d-1)}-1$, which is exponential in the tree depth $d$.

A feature of the tree graph is that leaves in the same layer are equivalent  to each other, called partial translation symmetry. Thus, specific types of TDFs can be reused to generate entanglement in the same layer. When the tree depth increasing, $(a-1)$ additional TDFs are required at each layer. As shown in Figs.~\ref{FIG3}(b) and (e), now only four TDFs are needed to generate a ${\rm{TCS}}_{2,4}$, much fewer than the case in Fig.~\ref{FIG3}(a). This optimization can be applied to arbitrary TCSs, and the exact number of TDFs required for generating ${\rm{TCS}}_{a,d}$ is $(d-1)(a-1)+(a-1)$, which is linear in the tree depth now.

How can a TDF be fully used in the generation process of a cluster state? One answer is to exploit the translational symmetry of the entanglement structure. For example, nodes in a 2D lattice structure are arranged at the same interval if they are in the same orthogonal direction. Thus, the generation process of a 2D lattice cluster state can entirely share the CZ gates induced by one TDF~\cite{PNAS_pichler_2017}. From this perspective, by projecting a graph representation of a cluster state into several orthogonal directions, it is possible to reuse TDF.

\subsection{Further Optimization of Binary TCSs}

This further optimization relies on the matrix representation to consider the virtual nodes in the generation process, that is, $|0\rangle$-state diagonal elements. We create a usable excitation space in advance, where the nodes represent available photon excitations, and the parallel edges are entanglement that can be induced by the same TDF. The nodes and edges in the available excitation space can be arbitrarily selected to form the desired entanglement structure. The fewer edge directions included in this construction, the fewer TDFs are required to generate the corresponding cluster state.

For brevity, we demonstrate the mapping from the ${\rm{TCS}}_{2,4}$ to a 2D lattice structure in Fig.~\ref{FIG3}(c). We stress that a 15-qubit TCS can be completely included within a 31-node 2D lattice structure. The additional 16 nodes are those that can be excited but have not been excited yet, defined as virtual nodes. Its corresponding matrix representation is shown in Fig.~\ref{FIG3}(f), meaning that we can generate ${\rm{TCS}}_{2,4}$ using only one additional TDF. The generation process can be written in the form
\begin{align}
    {\bf M}_{{\rm{TCS}}_{2,4}}=\prod_{n\in T'_{2}}E_{7,n} \prod_{m\in T'_{1}}E_{1,m} \prod_{i\in T'_0} X_{i} {\bf M}_0,
\end{align}
where $T'_{0}=\left\{1,7,8,11,13,...,19,21,24,25,31 \right\}$, $T'_{1}=\left\{7,13,...,18,24\right\}$, and $T'_{2}=\left\{1,8,11,14,17,24 \right\}$. 

\begin{figure}[t]
\centering
\includegraphics[width=0.95\linewidth]{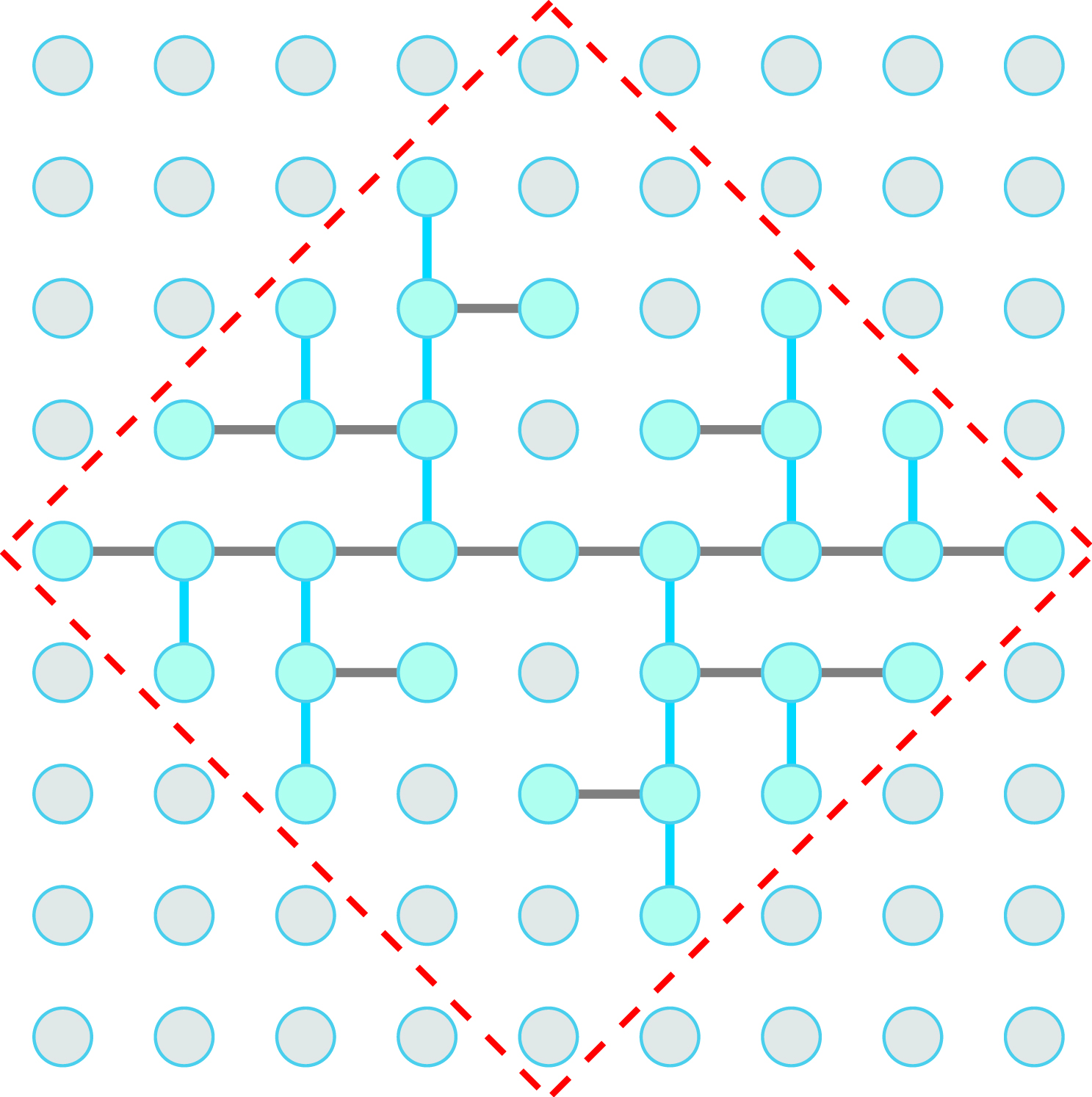}
\caption{Schematic diagram for mapping a TCS$_{2,5}$ into a 2D lattice structure. The size of the area available for the binary tree depends on its tree depth $d$. The longest connection between nodes in the binary tree forms the diagonal line of the available area. If the depth $d\ge 6$, an overlap of nodes will inevitably occur. }
\label{FIG4}
\end{figure}

Utilizing this mapping method and virtual nodes, cluster states with a binary tree structure can be easily generated, but there is a limitation on the size of the tree structure. In the generation of photonic cluster states, no more than one photon can exist at the same time. The number of nodes in the tree graph grows exponentially as the number of layers increases, but the growth of the number of nodes is just polynomial in the lattice structure. When increasing the number of layers, the lattice structure is unable to accommodate all nodes in the tree after a threshold. For example, the number of nodes in a binary tree is $N_{T}=2^d-1$, and the number of available nodes in the area it occupies is $N_{A}=d^2+(d-1)^2$. Comparing these two expressions, the threshold for $N_{A}>N_{T}$ is $d\leq5$, as shown in Fig.~\ref{FIG4}, meaning that with one additional TDF, we can at most generate ${\rm{TCS}}_{2,5}$.

If considering a tree with more branches, i.e., the case where $a>2$, the $a$-dimensional lattice structure is needed for mapping a tree graph. To facilitate a comparison of the vertex number in these two graphs, we define a function,
\begin{equation}
    F(a,d) = N_S - N_T = d^a + (d-1)^a - \left[\frac{(a^d-1)}{(a-1)} \right],
\end{equation}
where $a$ and $d$ are positive integers. Given the value of $a$, we can search for the maximum value $d_{\rm{max}}$ of $d$ that holds $F(a,d_{\rm{max}})>0$ as the threshold for generating the TCS$_{a,d}$ with additional $(a-1)$ kinds of TDFs. As $a$ increases, the number of layers $d$, obtained by this method, will approach $a$ until $d=a$.

So far, the number of TDFs required to generate cluster states can be efficiently estimated and optimized with the matrix representation. As in dealing with an acyclic graph, like binary trees, the number of TDFs required can be decreased from $\mathcal{O}(2^d)$ to $\mathcal{O}(d)$. Furthermore, after adding virtual nodes to map the tree structure with $a$ branches into an $a$-dimensional lattice structure, one can generate a finite size TCS$_{a,d}$ with just $(a-1)$ additional TDFs. When we only focus on the tree depth $d$, the requirement of TDFs is constant.

Constructing an excitation space with some orthogonal bases and mapping the desired entanglement structure into this space is a universal optimization method that can work for generating arbitrary cluster states. However, this method has little effect on annular entanglement structures, such as the complete graph (corresponding to CCSs). Overall, the optimization method enabled by the matrix representation allows for the efficient construction of diverse cluster states, thereby facilitating their practical implementation.


\section{Discussion}

When a series of TDFs have been introduced, photons are successively transmitted through different channels, and they may suffer additional loss during this process. Benefitting from the compact structure of our protocol, photons do not have to travel such long distances that their transmission losses are far from noticeable. Moreover, when guiding photons from one channel to another using circulators or routers, some of the photons may be leaked into the environment with probability $\gamma$.

The loss between channels can be described as amplitude damping $\mathcal{A}$ of photons~\cite{nielsen2010quantum}. Taking the photon absence ($|0\rangle$) and presence ($|1\rangle$) as the coding basis, amplitude damping turns each photon from $|1\rangle$ to $|0\rangle$ with probability $\gamma$. Tracing over the environment, this quantum operation on a single qubit can be written as
\begin{equation}
    \mathcal{E}(\rho)=K_{0}\rho K_{0}^{\dagger}+K_{1}\rho K_{1}^{\dagger}.
\end{equation}
Here $\rho$ is the density matrix of the initial state, $K_{0}$ and $K_{1}$ are Kraus operators, satisfying $\sum_{i}K_{i}^{\dagger}K_{i}=I$. In the amplitude-damping operation, their forms are
\begin{equation}
    K_{0}=
    \begin{pmatrix}
        1 & 0 \\
        0 & \sqrt{1-\gamma}
    \end{pmatrix},\quad
    K_{1}=
    \begin{pmatrix}
        0 & \sqrt{\gamma} \\
        0 & 0
    \end{pmatrix}.
\end{equation}
The fidelity of the state $\rho$ and operation $\mathcal{E}(\rho)$ is defined as
\begin{equation}
F\left(\rho,\mathcal{E}(\rho)\right)=\left(\text{Tr}\sqrt{\sqrt{\rho}\mathcal{E}(\rho)\sqrt{\rho}}\right)^2.
\end{equation}
Assuming the initial state is $|\psi\rangle=|+\rangle$, we can calculate the fidelity after amplitude damping
\begin{equation}
    F_{\rm{single}}=(1+\sqrt{1-\gamma})/2. 
\end{equation}

Considering a cluster state with $N$ photons, the general Kraus operators $\widetilde{K}$ are composed of the tensor product of the individual Kraus operators,
\begin{equation}
\widetilde{K}_{i}=K_{i}^{1}\otimes K_{i}^{2}\otimes\ldots\otimes K_{i}^{N}\quad \left(i\in\left\{0,1\right\}\right).
\end{equation}
It looks complicated, however, when we expand it and calculate the fidelity, the cross terms will cancel each other out because of the localization of the damping process. The fidelity of the $N$-photon cluster state after one amplitude damping operation $\mathcal{A}$ can be expressed by the product of single-photon contributions as
\begin{equation}
    F_{\mathcal{A}}=\left(F_{\rm{single}}\right)^{N}
    =\left(\frac{1+\sqrt{1-\gamma}}{2}\right)^{N}.
\end{equation}

Considering the amplitude damping between two channels (TDFs), the fidelity of cluster states decreases exponentially when increasing the number of qubits $N$. If there are multiple TDFs in the generation process, the amplitude damping operations will be repeated on each photon; thus, the fidelity decay rate will be multiplied. This feature reminds us that when using multiple TDFs to generate complex cluster states, the loss caused by additional channels is not negligible.

In addition to the amplitude damping operation, the imperfection of quantum gates during the generation process will also reduce the fidelity of the resulting cluster state. More precisely, the single-qubit gate (Hadamard gate) and the two-qubit gate (CZ gate) have different behaviors, and we will evaluate the effect of them respectively and discuss the total effect of these operations. 

The imperfection of the single-qubit gate can be simply described with the average fidelity $F_{S}$. In the generation of a cluster state, each qubit should be applied to a Hadamard gate ($\rm{H}$) first, and their imperfection will accumulate multiplicatively. The number of qubits contained in the cluster state can be extracted by tracing the operations of the distribution matrix
\begin{equation}
N_{\rm{H}}=\rm{Tr}({\bf D}),
\end{equation}
then the contribution of $\rm{H}$ gates to the overall fidelity is the $\left(F_{S}\right)^{N_{\rm{H}}}$ term. Recent advancements in photonic systems have enabled the fidelity of single-qubit gates to reach 99.9\% or even better.

As for the two-qubit gate, such as the CZ gate we applied in the generation of cluster states, the process fidelity $F_{T}$ is suitable for describing their effect on the fidelity of the final state. The number of CZ gates (denoted by $N_{\rm{CZ}}$) required depends on the entanglement structure of the cluster state, and the fidelity reduction caused by CZ gates can be represented as $\left(F_{T}\right)^{N_{\rm{CZ}}}$. If we have defined the representation matrix of the cluster state, we can get the relation directly,
\begin{equation}
N_{\rm{CZ}}=||{\bf D}||-\rm{Tr}({\bf D}).
\end{equation}
The process fidelity of the CZ gate in the photonic system has been improved to 99.6\%. It has a decisive effect on producing large-scale cluster states with high fidelity.

The number of amplitude damping operations $N_{\mathcal{A}}$ is one less than the number of TDFs, which can also be extracted from the matrix representation. In Appendix~\ref{imp}, we present a circulator structure for the single-photon case based on the nanophotonic system. In that case, the loss factor $\gamma$ depends on the dissipation factor of the chiral scattering process. Based on relevant literature, a reference value is $\gamma=0.98$.

Finally, the formula of the fidelity $F_{\mathcal{C}}$ of the resulting cluster state considering the imperfect gate operations and amplitude damping operations has the form
\begin{equation}
\label{fidelity}
F_{\mathcal{C}}=\left(F_{S}\right)^{N_{\rm{H}}}\cdot \left(F_{T}\right)^{N_{\rm{CZ}}} \cdot \left(F_{\mathcal{A}}\right)^{N_{\mathcal{A}}}.
\end{equation}

\begin{table}[t]
    \centering
    \renewcommand{\arraystretch}{1.3}
    \begin{tabular}{>{\centering\arraybackslash}m{1.6cm}|>{\centering\arraybackslash}m{2.1cm}>{\centering\arraybackslash}m{2cm}>{\centering\arraybackslash}m{2.1cm}}
    \hline
        \multirow{2}{10pt}{$N$} & ${\rm{TCS}}_{2,d}$ & ${\rm{TCS}}_{2,d}$ & \multirow{2}{22pt}{${\rm{CCS}}_{N}$} \\
        ~ & (Initial) & (Optimized) & ~ \\ \hline
        1 $(d=1)$ & $0.9990$ & $0.9990$ & $0.9990$\\
        3 $(d=2)$ & $0.9890$ & $0.9890$ & $0.9271$\\
        7 $(d=3)$ & $0.7306$ & $0.9694$ & $0.4501$\\
        15 $(d=4)$ & $0.1512$& $0.9314$ & $0.0126$\\
        31 $(d=5)$ & $1.338\times 10^{-4}$& $0.8596$ & $1.947\times 10^{-9}$\\ \hline
    \end{tabular}
    \caption{Fidelities of TCSs and CCSs}
    \label{FC}
\end{table}

Choosing ${\rm{TCS}}_{2,d}$ as examples, we can estimate the improvements that result from the optimization process based on matrix representation. First, the initial method requires $(2^{d-1}-1)$ kinds of TDFs for generating a ${\rm{TCS}}_{2,d}$; which means that each qubit would suffer $(2^{d-1}-2)$, where $d\ge 2$, amplitude damping operations. However, the optimized method in section~\ref{TCS} just needs one additional TDF to generate a ${\rm{TCS}}_{2,d}~(d\leq 5)$, which will not introduce any amplitude damping.

In the optimized method, the operations required for generating the $
{\rm{TCS}}_{2,d}$ can be enumerated in the form of ${\rm{H}}^{N}{\rm{CZ}}^{N-1}$, where $N$ is the total number of qubits. Note that it is the same as the requirement for the generation of a 1D cluster state with $N$ qubits. If using the initial method, the operations we need for generation have the form of ${\rm{H}}^{N}{\rm{CZ}}^{N-1}{\mathcal{A}}^{N\times\mathcal{O}(2^d)}$. The amplitude damping $\mathcal{A}$ results in additional dissipation that increases exponentially with the number of layers $d$. For comparison, the operations for generating CCSs containing the same number of qubits have the form ${\rm{H}}^{N}{\rm{CZ}}^{N(N-1)/2}{\mathcal{A}}^{N(N-2)}$.

Using Eq.~\eqref{fidelity} and selecting the parameter $F_{S}=99.9\%$, $F_{T}=99.6\%$ and $\gamma=0.98$, we calculate the fidelities of ${\rm{CCS}}_{N}$ and ${\rm{TCS}}_{2,d},(1\leq d\leq 5)$, including the optimized case and the initial case, which are shown in Table~\ref{FC}. It is obvious that the fidelity of the ${\rm{TCS}}_{2,d}$ prepared using the optimized method is significantly improved, which demonstrates that amplitude damping is a great obstacle for generating large-scale cluster states. 

In conclusion, the imperfect quantum gates and the amplitude damping between channels will affect the fidelity of the resulting cluster states in real physical systems. We discuss the mechanism of how these processes affect fidelity and provide a formula for calculating the final fidelity $F_{\mathcal{C}}$ of the resulting cluster state. It is worth noting that the amplitude damping $\mathcal{A}$, which was not considered in previous studies, can greatly affect the fidelity of the cluster states, and our optimization can significantly improve this influence.


\section{Conclusion}

In this paper, we present a universal protocol for efficiently generating arbitrary cluster states via multiple time-delayed feedbacks (TDFs), enabled by a novel matrix representation tailored to the characteristics of entanglement induced by TDFs. The matrix representation allows systematic optimization of the generation process, significantly reducing the requirement of TDFs. Notably, this protocol allows us to generate binary tree cluster states (${\rm{TCS}}_{2,d}$) with only one TDF. Finally, by quantifying fidelities of resulting states under realistic conditions, we highlight the feasibility of high-fidelity cluster state generation even in imperfect systems. Our research provides a scalable, resource-efficient framework for generating complex entanglement structures. The optimized method confirms the physical feasibility of our protocol and is a possible option for the preparation of photonic cluster states that can be used for MBQC, quantum communication, or quantum error correction.


\section{Acknowledgments.}

This work is supported by the National Natural Science Foundation of China (Grant No. 12375025), the National Key R\&D Program of China (Grant No. 2019YFA0308200). F.N. is supported in part by: the Japan Science and Technology Agency (JST) [via the CREST Quantum Frontiers program Grant No. JPMJCR24I2, the Quantum Leap Flagship Program (Q-LEAP), and the Moonshot R\&D Grant Number JPMJMS2061].



\appendix
\setcounter{figure}{0}
\renewcommand{\thefigure}{A\arabic{figure}}
\section{Physical implementation of the circulator}
\label{imp}

\begin{figure}[b]
\centering
\includegraphics[width=1\linewidth]{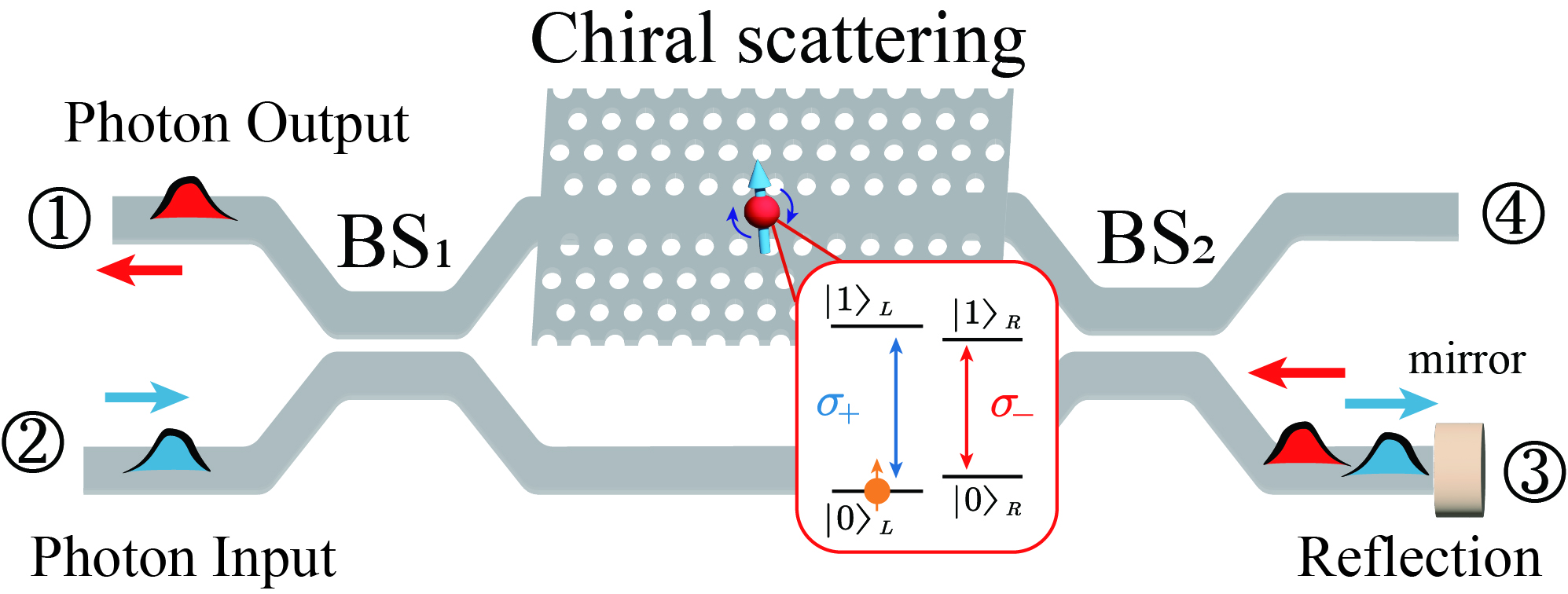}
\caption{Physical implementation of a photonic circulator. The colors of the wave packets correspond to the polarization, and port \ding{194} with a mirror outside will reflect photons. The emitter in the upper arm of MZI is set for the chiral scattering of photons.}
\label{FIG_A1}
\end{figure}

In our protocol, a circulator that can work in the single-photon regime could play an important role in the connection of different blocks. However, conventional optical circulators are mostly built on fiber or optical cavity platforms~\cite{Science_Michael_2016_circulator, LPL_wang_2020}, and they do not provide assurance in this regard. As a reference, we propose a physical implementation of the circulator in the nano-photonic system.

As shown in Fig.~\ref{FIG_A1}, the essential elements of our proposed circulator are the Mach-Zehnder interferometer (MZI)~\cite{Phys.Rep._Ma_2011,Phys.Rep._Oleh_2023} and the photon-emitter interface~\cite{PRA_Nori_2014,RevModPhys_lodahl_2015,NatNano_sollner_2015,PRX_borregaard_2020,LPL_wang_2020,NatNano_uppu_2021}. The emitter in the circulator initially has a V-type energy level consisting of one ground state ($|0\rangle$) and two excited states ($|1\rangle_L, |1\rangle_R$). In order to eliminate the effect of the reflected photons, an additional magnetic field is applied to split the ground state into $|0\rangle_{R}$ and $|0\rangle_{L}$ (depicted in the red box of Fig.~\ref{FIG_A1}). Only the right-propagating photons with right-handed circular polarization drive the $|0\rangle_R$ to $|1\rangle_R$ transition ($\sigma_{+}$), while other photons are decoupled from the emitter. 

This photon-emitter interface works as a chiral scattering~\cite{Phys.Rep._Konstantin_2015,Science_Konstantin_2015,PRA_Cai_2025,Nat.Pho._Zhang_2025,L-P-review_Huang_2022} in the upper arm of the MZI. When a right-propagating photon is incident on this interface, the coupling between the $\sigma_{+}$ transition causes it to undergo a $\pi$-phase scattering without reflection.

Back to the MZI, we assume that all the incident photons have right-handed circular polarization and are incident from port \ding{193}. The chirality of scattering will make them suffer a $\pi$-phase scattering in the upper arm. Based on the characteristics of MZI, the ratio of output from port \ding{194} can be calculated as $(1-\cos{\theta})/2$. If $\theta=\pi$, photons will only exit from port \ding{194}.

Afterwards, a mirror outside will reflect photons and reverse their polarization so that the reflected photons will enter the MZI from port \ding{194} with left-handed circular polarization. Because left-propagating photons do not interact with the emitter, it implies that photons will totally exit the MZI from port \ding{192}. This process [see Fig.~\ref{FIG_A1}], in fact, directs the reflected photons into another channel, thus realizing the function of a photonic circulator. 

The photonic circulator includes an MZI structure and a photon-emitter interface on one arm. The loss mainly occurs when (i) the mode is not purely circularly polarized at the position of the emitter and (ii) the emitter couples to modes other than the waveguide mode. 

We summarize these losses by a dissipation factor
\begin{equation}
    \beta = \dfrac{\Gamma_{R}}{\left(\Gamma_{\rm{wg}}+\gamma_{\rm{rad}}\right)},
\end{equation}
where $\Gamma_{\rm{wg}}=\Gamma_{R}+\Gamma_{L}$, $\Gamma_{R}$ and $\Gamma_{L}$ are the decay rates for coupling to the right and left propagating waveguide modes, respectively, and $\gamma_{rad}$ is the coupling to the non-guide radiation modes~\cite{NatNano_sollner_2015}. Considering the scattering process of photons with a single mode, the wave function after scattering can be written as
\begin{equation}
    |\Psi_{\rm{out}}\rangle = (1-2\beta)|\Psi_{\rm{in}}\rangle.
\end{equation}

This scattering result actually includes effects of a damped channel and a $\pi$-phase shift that serve the MZI. The evolution of the wave function in this chiral scattering process can be described by a set of Kraus operators:
\begin{equation}
    K^{c}_{0}=
     \begin{pmatrix}
        -\sqrt{2\beta-1} & 0 \\
        0 & 1
    \end{pmatrix},\quad
    K^{c}_{1}=
     \begin{pmatrix}
        0 & 0 \\
        \sqrt{2(1-\beta)} & 0
    \end{pmatrix}.
\end{equation}
Thus, we can extract the correspondence between $\beta$ and $\gamma$ as $\gamma=2(1-\beta)$, and a possible value can be selected as the reference in the discussion is $\gamma=0.98$.

\bibliography{reference}

\end{document}